\documentclass[aps,prl,preprint,showpacs,superscriptaddress,groupedaddress]{revtex4-1}
\usepackage{graphicx}
\usepackage{subfigure}
\usepackage{dcolumn}
\usepackage{bm}
\usepackage{amssymb}
\usepackage{color}

\begin{document}

\title{Coherent observations of gravitational radiation \\ with LISA and gLISA}
\author{Massimo Tinto}
\email{massimo.tinto@jpl.nasa.gov}
\affiliation{Jet Propulsion Laboratory,\\
California Institute of Technology, \\
4800 Oak Grove Drive, Pasadena, California, 91109, USA}

\author{Jos\'e C.N. de Araujo}
\email{jcarlos.dearaujo@inpe.br}
\affiliation{Divis\~ao de Astrof\'isica, Instituto Nacional de Pesquisas Espaciais \\
Avenida dos Astronautas 1758, S\~ao Jos\'e dos Campos, \\
S\~ao Paulo, 12227-010, Brasil 
}

\date{\today}

\begin{abstract}
  The geosynchronous Laser Interferometer Space Antenna (gLISA) is a
  space-based gravitational wave (GW) mission that, for the past five
  years, has been under joint study at the Jet Propulsion Laboratory,
  Stanford University, the National Institute for Space Research
  (I.N.P.E., Brazil), and Space Systems Loral. If flown at the same
  time as the LISA mission, the two arrays will deliver a joint
  sensitivity that accounts for the best performance of both missions
  in their respective parts of the milliHertz band.  This simultaneous
  operation will result in an optimally combined sensitivity curve
  that is “white” from about $3 \times 10^{-3} \ {\rm Hz}$ to $1 \
  {\rm Hz}$, making the two antennas capable of detecting, with high
  signal-to-noise ratios (SNRs), coalescing black-hole binaries (BHBs)
  with masses in the range $(10 - 10^8) M_\odot$. Their ability of
  jointly tracking, with enhanced SNR, signals similar to that
  observed by the Advanced Laser Interferometer Gravitational Wave
  Observatory (aLIGO) on September 14, 2015 (the GW150914 event) will
  result in a larger number of observable small-mass binary
  black-holes and an improved precision of the parameters
  characterizing these sources. Together, LISA, gLISA and aLIGO will
  cover, with good sensitivity, the $(10^{-4} - 10^3) \ {\rm Hz}$
  frequency band.
\end{abstract}

\maketitle

The first direct observation of a GW signal announced by the LIGO
project \cite{LIGO} on February 11, 2016 \cite{GW150914}, represents
one of the most important achievements in experimental physics today,
and marks the beginning of GW astronomy \cite{Thorne1987}. By
simultaneously measuring and recording strain data with two
interferometers at Hanford (Washington) and Livingston (Louisiana),
scientists were able to reach an extremely high level of detection
confidence and infer unequivocally the GW source of the observed
signal to be a coalescing binary system containing two black-holes of
masses $M_1 = 36^{+5}_{-4} \ M_\odot$ and $M_2 = 29^{+4}_{-4} \
M_\odot$ out to a luminosity distance of $410^{+160}_{-180} \ Mpc$
corresponding to a red-shift $z = 0.09^{+0.03}_{-0.04}$ (the above
uncertainties being at the $90$ percent confidence level). The network
of the two LIGO interferometers could constrain the direction to the
binary system only to a broad region of the sky because the
French-Italian VIRGO detector \cite{VIRGO} was not operational at the
time of the detection and no electromagnetic counterparts could be
uniquely associated with the observed signal.

Ground-based observations inherently require use of multiple detectors
widely separated on Earth and operating in coincidence. This is
because a network of GW interferometers operating at the same time can
(i) very effectively discriminate a GW signal from random noise and
(ii) provide enough information for reconstructing the parameters
characterizing the wave's astrophysical source (such as its
sky-location, luminosity distance, mass(es), dynamical time scale,
etc.)  \cite{SchutzTinto,GT}. Space-based interferometers instead,
with their six links along their three-arms, have enough data
redundancy to validate their measurements and uniquely reconstruct an
observed signal \cite{PPA1998}. A mission such as LISA, with its three
operating arms, will be able to assess its measurements' noise level
and statistical properties over its entire observational frequency
band. By relying on a time-delay interferometric (TDI) measurement
that is insensitive to gravitational waves \cite{TAE}, LISA will
assess its in-flight noise characteristics in the lower part of the
band, i.e. at frequencies smaller than the inverse of the
round-trip-light time. At higher frequencies instead, where it can
synthesize three independent interferometric measurements, it will be
able to perform a data consistency test by relying on the null-stream
technique \cite{GT, TL, SST}, i.e. a parametric non-linear combination
of the TDI measurements that achieves a pronounced minimum at a unique
point in the search parameter space when a signal is present.  In
addition, by taking advantage of the Doppler and amplitude modulations
introduced by the motion of the array around the Sun on long-lived GW
signals, LISA will measure the values of the parameters associated
with the GW source of the observed signal \cite{PPA1998}.

Although a single space-based array such as LISA can synthesize the
equivalent of four interferometric (TDI) combinations (such as the
Sagnac TDI combinations $(\alpha, \beta, \gamma, \zeta)$)
\cite{TD2014}, its best sensitivity level is achieved only over a
relatively narrow region of the mHz frequency band. At frequencies
lower than the inverse of the round-trip light time, the sensitivity
of a space-based GW interferometer is determined by the level of
residual acceleration noise associated with the nearly free-floating
proof-masses of the onboard gravitational reference sensor and the
size of the arm-length. This is because in this region of the
accessible frequency band the magnitude of a gravitational wave signal
in the interferometer's data scales linearly with the arm-length. At
frequencies higher than the inverse of the round-trip light time
instead, the sensitivity is primarily determined by the
photon-counting statistics \cite{TAAA}. Note that this noise grows
linearly with the arm-length through its inverse proportionality to
the square-root of the received optical power and because, in this
region of the accessible frequency band, the gravitational wave signal
no-longer scales with the arm-length. In summary, for a given
performance of the onboard science instrument and optical
configuration, the best sensitivity level and the corresponding
wideness of the bandwidth over which it is reached are uniquely
determined by the size of the array.

The best sensitivity level and the corresponding bandwidth over which
it is reached by a space-based interferometer become particularly
important when considering signals sweeping up-wards in frequency such
as those emitted by coalescing binary systems containing
black-holes. As recently pointed out by Sesana \cite{Sesana}, it is
theoretically expected that a very large ensemble of coalescing binary
systems, with masses comparable to those of GW150914, will have
characteristic wave's amplitudes that could be observable by LISA over
an accessible frequency region from about $1.5 \times 10^{-2}$ Hz to
about $7.6 \times 10^{-2}$ Hz. These frequency limits correspond to
(i) the assumption of observing a GW150914-like signal for a period of
five years (approximately equal to its coalescing time), and (ii) the
value at which the signal's amplitude is equal to the LISA
sensitivity. Although one could in principle increase the size of the
LISA's optical telescopes and rely on more powerful lasers so as to
increase the upper frequency cut-off to enlarge the observational
bandwidth, in practice pointing accuracy and stability requirements
together with the finiteness of the onboard available power would
result in a negligible gain.

A natural way to broaden the mHz band, so as to fill the gap between
the region accessible by LISA and that by the ground interferometers,
is to simultaneously fly an additional interferometer with a smaller
arm-length. The geosynchronous Laser Interferometer Space Antenna
(gLISA) \cite{TAAA,TBDT,TAKAA}, which has been analyzed during the
past five years by a collaboration of scientists and engineers at the
Jet Propulsion Laboratory, Stanford University, the National Institute
for Space Research (I.N.P.E., Brazil), and Space Systems Loral, would
allow this \cite{footnote1}.  With an arm-length of about $7.3 \times
10^4$ km, it could achieve a shot-noise limited sensitivity in the
higher region of its accessible frequency band that is about a factor
of $70$ better than that of LISA \cite{footnote2}; gLISA will display
a minimum of its sensitivity in a frequency region that perfectly
complements those of LISA and aLIGO, resulting into an overall
accessible GW frequency band equal to $(10^{-4} - 10^3) \ {\rm Hz}$.

To derive the expression of the joint LISA-gLISA sensitivity, we first
note that the noises in the TDI measurements made by the two arrays
will be independent. This means that the joint signal-to-noise ratio
squared, averaged over an ensemble of signals randomly distributed
over the celestial sphere and random polarization states, $\langle
SNR_{eL + gL}^2 \rangle$, can be written in the following form
\begin{eqnarray}
\langle SNR_{eL +  gL}^2 \rangle & \equiv & 4 \int_{f_1}^{f_2}
\frac{|{\widetilde h} (f)|^2}{S_h^{eL + gL} (f)} \ df \  
= \langle SNR_{eL}^2 \rangle + \langle SNR_{gL}^2 \rangle
\nonumber
\\
& = & 4 \int_{f_1}^{f_2} \left\{ \frac{1}{S_h^{eL} (f)} +
      \frac{1}{S_h^{gL} (f)} \right\}
   |{\widetilde h} (f)|^2 \ df \ ,
\label{SNRnet}
\end{eqnarray}
where the functions $S_h^i (f), \ i = eL \ , \ gL$ correspond to the
squared optimal sensitivities (as defined through the $A$, $E$, and
$T$ modes \cite{PTLA, TD2014}) of the LISA and gLISA arrays
respectively. Note that the factor $4$ in front of the integral in
Eq.(\ref{SNRnet}) reflects the adopted convention throughout this
article of using one-sided power spectral densities of the noises
\cite{FH1998,TA}.

\begin{figure}
\includegraphics[width = \linewidth, clip]{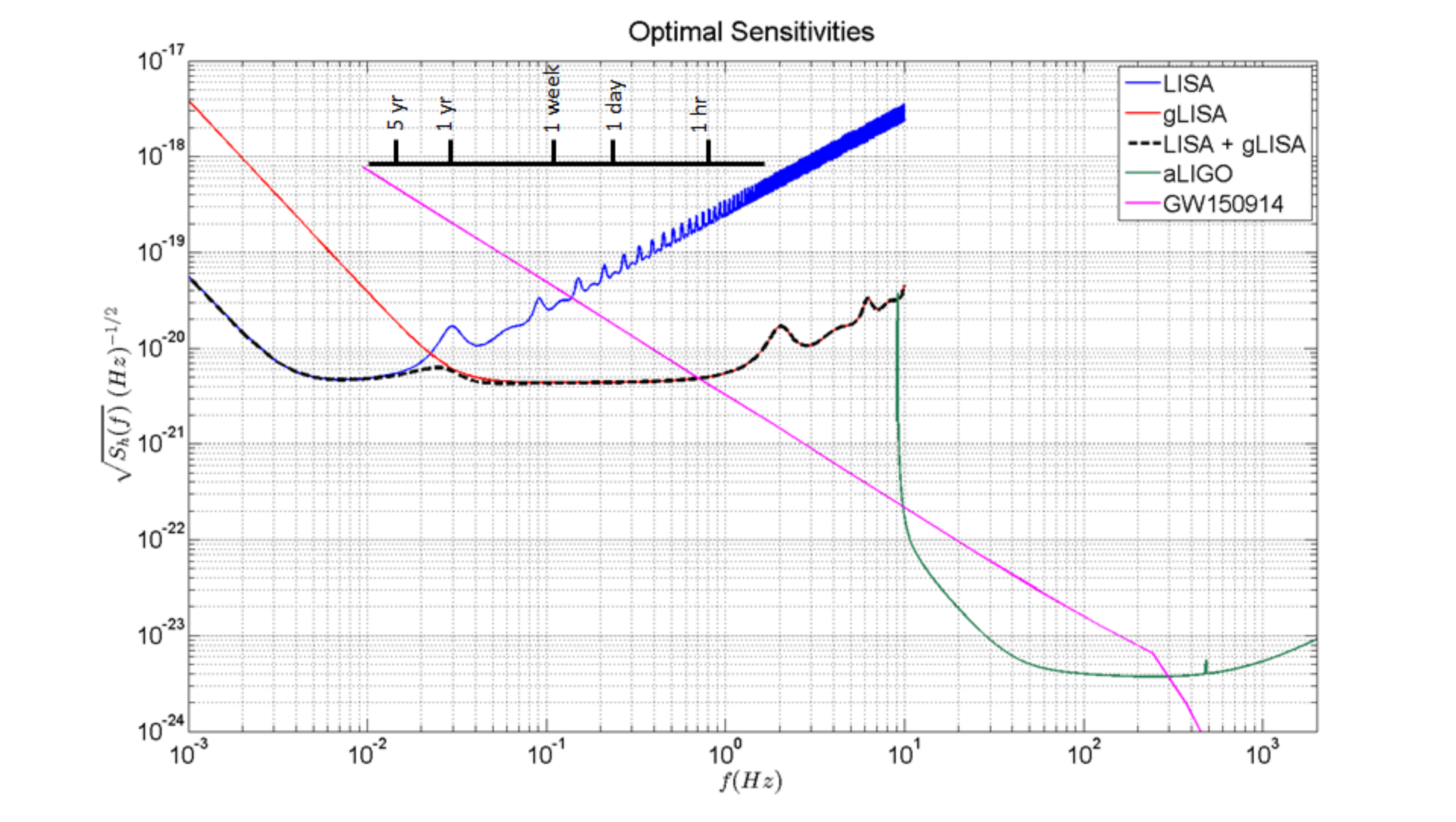}
\caption{The optimal sensitivities of LISA (blue) and gLISA (red)
  averaged over sources randomly distributed over the sky and
  polarization states. Each of these curves is uniquely determined by
  the TDI $A$, $E$, and $T$ combinations associated with the data from
  each mission. (see \cite{PTLA} for the derivation of the $A$, $E$,
  and $T$ modes and the resulting optimal sensitivity. Also refer to
  \cite{TD2014} for the revised expressions of their noise
  spectra). The dashed-line (black) is the result of optimally
  combining the sensitivities of LISA and gLISA. As expected, in the
  lower part of the mHz band, LISA defines the network sensitivity; in
  the overlapping region the two sensitivities smoothly blend,
  resulting in a maximum sensitivity gain of $\sqrt{2}$. In the higher
  part of the accessible band gLISA defines the joint LISA-gLISA
  sensitivity. For completeness we have included the projected aLIGO
  sensitivity (green) together with the amplitude of GW150914
  (magenta) as a function of the Fourier frequency $f$.  Also included
  is a black horizontal line showing the frequency evolution of
  GW150914 over a 5 years period before coalescence.}
\label{fig1}
\end{figure}

From Eq. (\ref{SNRnet}) it is straightforward to derive the following
expression of the joint sensitivity squared, $S_h^{eL + gL} (f)$, in
terms of the individual ones, $S_h^{eL} (f)$ and $S_h^{gL} (f)$
\begin{equation}
  S_h^{eL + gL} (f) \equiv \frac{S_h^{eL} (f) \ S_h^{gL} (f)}{S_h^{eL} (f) + S_h^{gL} (f)} \ .
\label{network}
\end{equation}
In Fig. \ref{fig1} we plot the optimal sensitivities of LISA (blue),
gLISA (red) averaged over sources randomly distributed over the sky
and polarization states, together with the joint LISA-gLISA (black)
given by Eq. (\ref{network}). For completeness, and to visually
exemplify the scientific advantages of flying simultaneously two
space-based missions of different arm-lengths, we have included the
anticipated sensitivity of the third-generation aLIGO detector (green)
together with the amplitude of the gravitational wave signal GW150914
(magenta) as functions of the Fourier frequency $f$. We have also
included a black horizontal line showing the frequency evolution of
GW150914 over a 5 years period before coalescence.

Shortly before the announcement made by aLIGO of the detection of a
second signal emitted by another black-hole binary system, GW151226,
with masses roughly half of those of GW150914 \cite{GW150914,
  GW151226}, Sesana estimated that a large number of such systems
\cite{Sesana} could be observed by LISA while they are still spiraling
around each other over periods as long as the entire five years
duration of the mission. Because of these compelling astrophysical
reasons, to quantify the scientific advantages of flying gLISA jointly
with LISA we will focus our attention on signals emitted by coalescing
black-hole binaries with chirp masses in the range $(10 - 100)
M_\odot$.

In our analysis we will use the following expressions (valid for
circular orbits) of the Fourier transform of the amplitude of the
gravitational wave signal emitted by such systems, $\widetilde{h}
(f)$, and the time, $t_c$, it takes them to coalesce \cite{Maggiore}
\begin{eqnarray}
\widetilde{h} (f) & = & \sqrt{\frac{5}{24}} \ \pi^{-2/3} \
                        \left[\frac{(1 + z) G M_c}{c^3}\right]^{5/6} \
                        \frac{c}{D_L} \ f^{-7/6} \ ,
\label{h}
\\
t_c & = & \frac{5}{256} \ \left[\frac{(1 + z) G M_c}{c^3}\right]^{-5/3} \ (\pi f_{gw})^{-8/3} \ .
\label{tc}
\end{eqnarray}
In Eqs.(\ref{h}, \ref{tc}) $G$ is the gravitational constant, $c$ is
the speed of light, $z$ is the cosmological red-shift, $D_{L}$ is the
corresponding luminosity distance, $M_c \equiv (M_1 M_2)^{3/5}/(M_1 +
M_2)^{1/5}$ is the chirp mass associated with the binary system whose
components have masses $M_1$ and $M_2$, $f$ is the Fourier frequency
and $f_{gw}$ the instantaneous frequency of the emitted GW.

By using the above signal amplitude and time to coalescence, and by
further assuming an integration time of $5$ years (which, in the case
of the GW150914 signal, defines the lower-limit of integration in the
integral of the SNR to be equal to $0.0151$ Hz), we have estimated the
SNRs achievable by the two interferometers when operating either as
stand-alone or jointly. We find LISA can observe a GW150914-like
signal with a SNR equal to $10.7$, while gLISA with an SNR of $14.4$
because of its better sensitivity over a larger part of its observable
band (see Fig. \ref{fig1}). As expected, the joint LISA-gLISA network
further improves upon the SNR of gLISA-alone by reaching a value of
about $18.0$. From these results we can further infer that gLISA will
achieve a sufficiently high SNR to warrant the detection of a
GW150914-like signal by integrating for a shorter time. We find that,
by integrating for a period of $135$ days prior to the moment of
coalesce of a GW150914-like system, gLISA can achieve a SNR of $10.7$.

The level of SNR achievable by the LISA-gLISA network over a five-year
integration is about $80$ percent higher than that of LISA alone. This
implies, from the estimated parameter precisions derived by Sesana
\cite{Sesana} and their dependence on the value of the SNR
\cite{CF1994}, that the LISA-gLISA network will estimate the
parameters associated with the GW source of the observed signal with a
precision that is $1.8$ better than that obtainable by LISA alone. It
should be said, however, that this is a lower bound on the improved
precision by which the parameters can be estimated since it is based
only on SNR considerations. As pointed out by McWilliams in an
unpublished document \cite{McWilliams}, the Doppler frequency shift
together with the larger diurnal amplitude modulations experienced by
the GW signal in the gLISA TDI measurements will further improve the
precision of the reconstructed parameters beyond that due to only the
enhanced SNR. We will analyze and quantify this point in a follow-up
article as this is beyond the scope of this letter.

\begin{figure}[htbp]
\includegraphics[width = \linewidth, clip]{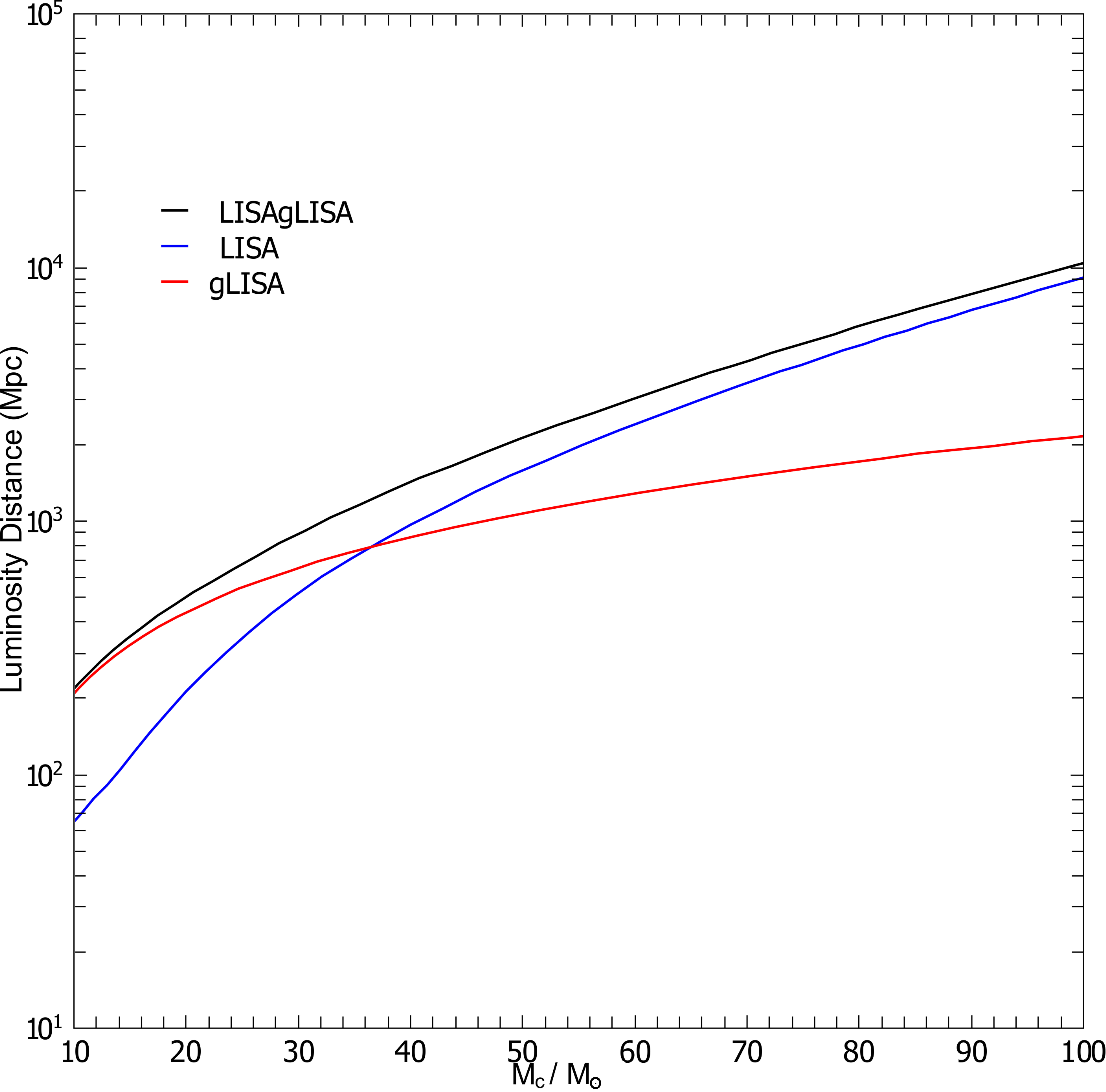}
\caption{Average luminosity distance achievable by LISA, gLISA and the
  LISA-gLISA network as function of the chirp mass. The SNR of each
  configuration has been assumed to be equal to $10$ with an
  integration time of five years.}
\label{fig2}
\end{figure}

From the expressions of the SNR and of the Fourier amplitude of the GW
signal, $\tilde{h}(f)$ (Eqs. \ref{SNRnet}, \ref{h}), and by fixing the
SNR to a specific value for each operational configuration
(stand-alone vs. network), it is possible to infer the corresponding
average luminosity distance to a BHB in terms of its chirp mass
parameter \cite{footnote3}. In Fig. \ref{fig2} we plot the results of
this analysis by assuming a SNR of $10$. As expected, the three
derived luminosity distances are monotonically increasing functions of
the chirp-mass. For chirp-masses in the range $[10 - 36) \ M_\odot$
gLISA can see signals further away than LISA because of its better
sensitivity at higher frequencies. Systems with a chirp-mass larger
that $36 \ M_\odot$ instead can be seen by LISA at a larger luminosity
distance than that achievable by gLISA alone. Note that the LISA-gLISA
network out-performs the stand-alone configurations by as much as $40$
percent for BHBs with chirp-masses in the interval $(30 - 40) \
M_\odot$. This results into a number of observable events that is
about $3$ times larger than that detectable by each interferometer
alone. Finally, a GW150914-like signal characterized by a chirp mass
of $28.1 \ M_\odot$ can be see by LISA at an average luminosity
distance of about $400 \ Mpc$, by gLISA out to $600 \ Mpc$, and by the
LISA-gLISA system out to about $800 \ Mpc$.

\begin{figure}[htbp]
\includegraphics[width = \linewidth, clip]{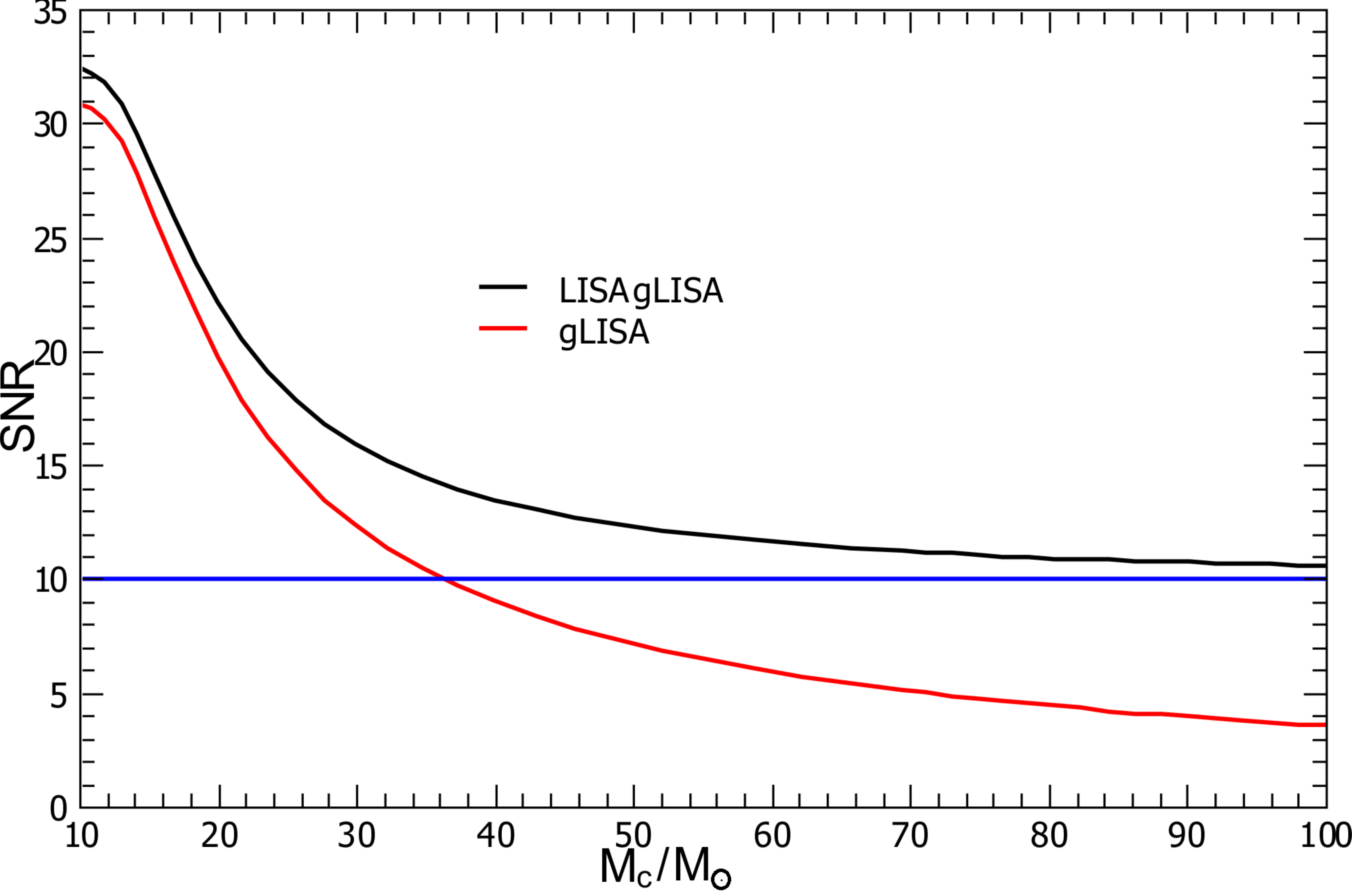}
\caption{SNR for gLISA and the LISA-gLISA network as a function of
  the chirp-mass of BHBs that are seen by LISA with an SNR = 10. An
  integration time of five years is also assumed.}
\label{fig3}
\end{figure}

In addition to derive the luminosity distance as a function of the
chirp-mass (with a chosen value of the SNR), we have also calculated
the SNR as a function of the chirp-mass by assuming the SNR of the
LISA mission to be constant and equal to $10$ (see
Fig. \ref{fig3}). This of course implies that the luminosity distance
of the system is uniquely determined by the chirp-mass through the
assumed SNR value of $10$ for LISA. Since we are focusing on
chirp-masses defined in the interval $[10 - 100] \ M_\odot$, from
Fig. \ref{fig2} we infer that the values assumed by the luminosity
distance will be greater than or equal to about $65 \ $ Mpc. This
minimum value of the luminosity distance defines an astrophysical
interesting region as it is more than four times larger than the
luminosity distance to the Virgo cluster of galaxies.

Fig. \ref{fig3} shows the two SNRs of gLISA and LISA-gLISA to be
decreasing functions of the chirp-mass, with their maximum values of
$31$ and $32.5$ respectively at $M_c = 10 \ M_\odot$. At $M_c \simeq
36 \ M_\odot$ instead the gLISA SNR also becomes equal to $10$ while,
as expected, the LISA-gLISA SNR is $\sqrt{2}$ larger.

This letter has shown the scientific advantages of simultaneously
flying with the LISA mission an additional, smaller size, space-based
interferometer such as gLISA. Because of its smaller arm-length, gLISA
will be able to survey the region of the GW frequency band that is in
between those accessible by LISA and aLIGO. By covering the entire mHz
and kHz GW frequency band, LISA, gLISA and aLIGO will detect all known
sources emitting in this broad frequency region, and observe signals
requiring multi-band detection for understanding the physical nature
of their sources.

\section*{Acknowledgments}

M.T. would like to thank Professor Daniel DeBra and Dr. Sasha Buchman
for many stimulating conversations about the gLISA mission concept,
Dr. John W. Armstrong for reading the manuscript and his valuable
comments, and Drs. Anthony Freeman and Daniel McCleese for for their
constant encouragement. M.T. also acknowledges financial support
through the Topic Research and Technology Development program of the
Jet Propulsion laboratory. J.C.N.A. acknowledges partial support from
FAPESP (2013/26258-4) and CNPq (308983/2013-0). This research was
performed at the Jet Propulsion Laboratory, California Institute of
Technology, under contract with the National Aeronautics and Space
Administration.


\begin{thebibliography}{99}
\bibitem{footnote2} The derivation of the gLISA sensitivity and the
  related discussion on the magnitude of the noises that define it
  have been presented in the main body and appendix of Ref.
  \cite{TAAA}. Although it was stated there that additional technology
  developments were needed to reduce the optical bench noise level
  below that of the photon-shot noise, the LISA Pathfinder experiment
  \cite{LPF} has demonstrated this noise to be a factor of $10^3$
  smaller than its anticipated value.
\bibitem{footnote3} To derive the redshift, $z$, we have assumed a
  fiducial $\Lambda$CDM flat cosmology with the matter density
  parameter $\Omega_{M} = 0.3$, the cosmological constant density
  parameter $\Omega_{\Lambda}=0.7$, and the value of the Hubble
  parameter at the present time $H_{0} = 70\, {\rm km\, s^{-1}\,
    Mpc^{-1}}$. To obtain the luminosity distance, $D_L$, in terms of
  the chirp mass, $M_c$, we have rewritten the SNR as a function of
  the redshift by expressing $D_{L}$ in terms of $z$ (see Eq. (13) in
  \cite{TAAA}). Then, for a given SNR, we have solved a transcendental
  equation to obtain z, and from it derived the luminosity distance.
\end{thebibliography}
\end{document}